# Formation of a stable surface oxide in MnBi$_2$Te$_4$ thin films

Golrokh Akhgar[1,2,3#], Qile Li[1,2,3#], Iolanda Di Bernardo[1,3], Chi Xuan Trang[1,3], Chang Liu[1,2,3], Julie Karel[2,3], Anton Tadich[4], Michael S. Fuhrer[1,3], Mark T. Edmonds[1,3*]

[1]School of Physics and Astronomy, Monash University, Clayton, VIC, Australia

[2]Department of Materials Science and Engineering, Monash University, Clayton, VIC, Australia

[3]ARC Centre for Future Low Energy Electronics Technologies, Monash University, Clayton, VIC, Australia

[4]Australian Synchrotron, 800 Blackburn Road, Clayton, VIC, Australia

# contributed equally

*Corresponding Author mark.edmonds@monash.edu

**ABSTRACT**

Understanding the air-stability of MnBi$_2$Te$_4$ thin films is crucial for the development and long-term operation of electronic devices based around magnetic topological insulators. In the present work, we study MnBi$_2$Te$_4$ thin films upon exposure to atmosphere using a combination of synchrotron-based photoelectron spectroscopy, room temperature electrical transport and atomic force microscopy to determine the oxidation process. After 2 days air exposure a 2 nm thick oxide passivates the surface, corresponding to oxidation of only the top two surface layers, with the underlying layers preserved. This protective oxide layer results in samples that still exhibit metallic conduction even after several days air exposure. Furthermore, the work function decreases from 4.4 eV for pristine MnBi$_2$Te$_4$ to 4.0 eV after the formation of the oxide, along with only a small shift in the core levels indicating minimal doping as a result of air exposure. With the oxide confined to the top surface layers, and the underlying layers preserved, it may be possible to explore new avenues in how to handle, prepare and passivate future MnBi$_2$Te$_4$ devices.

## INTRODUCTION

The intrinsic magnetic topological insulator, $MnBi_2Te_4$ is a bulk anti-ferromagnetic topological insulator that possesses both intrinsic magnetism and topological protection.[1-3] As a layered van der Waals material, $MnBi_2Te_4$ is similar to the well-known topological insulator $Bi_2Te_3$, with an additional Mn-Te layer inserted into the quintuple layer structure of $Bi_2Te_3$ to form a unit cell with a septuple layer (SL) configuration of Te-Bi-Te-Mn-Te-Bi-Te. The magnetic order comes from $Mn^{2+}$ ions, where the odd number of layers results in unpaired Mn spin and gives a net magnetic moment. The moments are coupled ferromagnetically within each SL and antiferromagnetically between adjacent SLs, forming A-type antiferromagnetic (AFM) states in the bulk of $MnBi_2Te_4$.[1-3]

As a layered material $MnBi_2Te_4$ can be mechanically exfoliated or grown via molecular beam epitaxy down to single layer or few-septuple layer thickness.[4-8] This allows observation of the thickness dependent transition from 2D ferromagnetic insulator in 1SL,[8] to axion and quantum anomalous Hall insulator phases in even and odd layer $MnBi_2Te_4$ respectively.[6, 7] Currently, $MnBi_2Te_4$ research has been focussed on minimizing exposure to atmosphere with careful preparation in inert or ultra-high vacuum conditions, followed by encapsulation or passivation with air-stable overlayers.[4-7] Little attention has been paid to the behaviour of pristine $MnBi_2Te_4$ thin films in air with the expectation that there is rapid oxidation of the surface. Despite this, there is relatively little understanding of the rate of oxidation or the thickness of the oxide on the surface.[9] Other 3D topological insulators, $Bi_2Te_3$ and $Bi_2Se_3$ that are structurally similar to $MnBi_2Te_4$ exhibit various degrees of reactivity, ranging from negligible surface reactivity [10] to a surface reconstruction forming bismuth bilayers upon exposure to air.[11] Crucially, the possibility of using a native oxide as a stable protective layer that protects the pristine layers of $MnBi_2Te_4$ underneath has yet to be explored experimentally. This lack of understanding of how $MnBi_2Te_4$ behaves in air necessitates a detailed study of the oxidation dynamics of $MnBi_2Te_4$ thin films.

In the present work, we utilise surface sensitive photoelectron spectroscopy, atomic force microscopy and room temperature electrical transport measurements on $MnBi_2Te_4$ thin films grown via molecular beam epitaxy (MBE) in order to study the oxidation process from brief exposure to atmosphere (1 hour) to prolonged exposure (2 to 14 days). We demonstrate that even after prolonged air exposure the oxidation remains confined to the top two septuple layers, leaving the deeper layers intact. The realization of a stable oxide confined to the surface of $MnBi_2Te_4$ provides new insight into the stability of this material, and may offer new pathways towards the development of electronic devices based around $MnBi_2Te_4$ that can be prepared or operate in ambient conditions.

## EXPERIMENTAL METHODS

### Growth of MnBi$_2$Te$_4$ thin films via molecular beam epitaxy

High-quality MnBi$_2$Te$_4$ thin films were grown by molecular beam epitaxy (MBE) in ultra-high vacuum (UHV). Growth on insulating sapphire ($\alpha$-Al$_2$O$_3$(0001)) was adopted in order to study the resistivity changes as a function of air exposure, whilst MnBi$_2$Te$_4$ grown on conductive Si(111) was used for the atomic force microscopy and X-ray photoelectron spectroscopy to understand the surface morphology and oxide chemistry. For all MnBi$_2$Te$_4$ film growths, effusion cells were used to evaporate elemental Bi (99.999%) and Mn (99.9%) in an overflux of Te (99.95%). Growth rates were calibrated with a quartz crystal microbalance.

Growth of MnBi$_2$Te$_4$ on $\alpha$-Al$_2$O$_3$(0001) utilises a two-step growth method, by modifying a recipe for ultra-thin Bi$_2$Te$_3$/Bi$_2$Se$_3$ or MnBi$_2$Se$_4$ grown on sapphire.[12] Atomically flat sapphire substrates (Shinkosha© Japan) were annealed in UHV at 600°C to outgas for an hour before annealing up to 800°C for 20 minutes to produce the (1×1) surface reconstruction of clean sapphire. An initial seeding layer of 1-quintuple layer (QL) of Bi$_2$Te$_3$ was grown at 150°C to aide nucleation. The temperature was then increased to 225°C, with another 1-QL of Bi$_2$Te$_3$ grown. A bilayer of MnTe was then deposited to allow for the spontaneous formation of a single SL of MnBi$_2$Te$_4$.[5] This was repeated 5 times, with each additional SL subjected to 5 minutes of post annealing in a Te overflux to improve crystallinity. To characterise the crystallinity of the film during growth reflection high-energy electron diffraction (RHEED) was performed as shown in the upper panel of Fig. 1b, with the sharp RHEED streaks indicating atomically flat single crystalline MnBi$_2$Te$_4$. The lower panel of Fig. 1b shows low energy electron diffraction and the expected (1×1) pattern corresponding to (111)-oriented growth. To prepare MnBi$_2$Te$_4$ films on Si(111), we used a previously established growth recipe that allows growth of MnBi$_2$Te$_4$ down to a single septuple layer thickness.[8] Briefly, the Si(111) was flash annealed to 1200 °C until the (7×7) surface of clean Si(111) was observed and the substrate temperature was then cooled and held at 230°C upon which the first QL Bi$_2$Te$_3$ was grown, followed by 1-BL of MnTe which spontaneously re-arranges into 1 SL of MnBi$_2$Te$_4$. This recipe was repeated until 5 SL MnBi$_2$Te$_4$ was grown, with RHEED used to monitor film quality during growth.

### Electrical Transport measurements

In order to perform transport electrical measurements after establishing the optimal growth conditions, MnBi$_2$Te$_4$ was then grown through a Ta shadow mask with Hall bar geometry onto a sapphire substrate that had pre-deposited Pd electrodes. This allowed growth of the MnBi$_2$Te$_4$ thin film to be grown in a Hall bar geometry without the need for additional patterning or fabrication steps as shown in Figure

1c. Figure 1d shows the optical image of the MnBi$_2$Te$_4$ Hall bar device. After growth on sapphire the MnBi$_2$Te$_4$ films were immediately transferred to an inter-connected Argon filled glovebox without exposure to air. Samples were mounted onto a chip carrier and wire bonded using indium inside the glovebox to minimise air exposure, then removed from the glovebox into air where resistivity measurement were started with less than 10 minutes exposure to air.

**Atomic force microscopy**

AFM measurements were performed using 70 kHz Si cantilevers using a Bruker Dimension Icon SPM in Tapping Mode at 256 samples per line and a scanning rate of 0.5 Hz.

**X-ray photoelectron spectroscopy**

For the XPS study, 5 SL MnBi$_2$Te$_4$ thin films were grown on Si(111) and then capped with a layer of Te and then transported in air to the Soft X-ray Beamline of the Australian Synchrotron and then introduced into ultra-high vacuum. The samples were annealed in the preparation chamber at 260°C for 2 hours to remove the capping layer, with LEED used to confirm the successful removal of the capping layer. Following the decapping, the sample was transferred under UHV to the analysis chamber where XPS measurements were performed using a SPECS Phoibos-150 Spectrometer at room temperature. The Bi 5$d$, Te 4$d$ and Mn 3$p$ core levels were measured at photon energies of 100 eV, 350 eV, 850 eV and 1486 eV. This ensured surface sensitivity for the low photon energy scans at 100 eV, with the higher photon energies used to characterize the depth dependence of the core levels, to confirm there was only MnBi$_2$Te$_4$ throughout the film. The MnBi$_2$Te$_4$ samples were exposed to atmosphere via venting the endstation's load-lock chamber with nitrogen gas and then exposing the sample to ambient conditions. The overall measurement uncertainty in the measured binding energies was ±0.02 eV. The work function was determined from the cut off in the secondary electron distribution with an experimental uncertainty of ±0.05 eV. The binding energy scale of all spectra are referenced to the Fermi energy ($E_F$), determined using either the Fermi edge or 4$f$ core level of an Au reference foil in electrical contact with the sample. Core level spectra were analyzed using a Shirley background subtraction, and then peak fitted using Voigt functions for each peak component.

**RESULTS & DISCUSSION**

We first study the room temperature transport properties of 5 SL as-grown and Te capped MnBi$_2$Te$_4$ films. Fig 1(e) shows sheet resistance, $R_{sheet}$, as a function of exposure time to air for uncapped and 20 nm Te capped (a standard method for protecting topological insulators in air) MnBi$_2$Te$_4$ devices. The sheet resistivity of the Te capped MnBi$_2$Te$_4$ (black line) remains at 9 kΩ/□ confirming it protects the MnBi$_2$Te$_4$ from ambient conditions, and consistent with previous reports.[5] We then measured uncapped MnBi$_2$Te$_4$ (red line), and surprisingly the sample does not show rapid degradation in air. At the start it possesses a resistivity of 14 kΩ/□, suggesting there is some initial degradation from the as-grown resistivity however, even after 50-60 hours the resistivity has only increased to 24 kΩ/□. This indicates that metallic transport is preserved, even though MnBi$_2$Te$_4$ may become oxidized when exposed to ambient conditions.

To understand the surface oxidation further and its effect on the surface morphology we performed atomic force microscopy (AFM) on uncapped 5 SL MnBi$_2$Te$_4$ grown on Si(111) as a function of time. After growth the sample was sealed in an argon filled container and transported to the AFM in order to minimise air exposure before the AFM measurements were taken. Fig 2(a) shows a large area topography image (800 nm x 800 nm), the film is predominantly a single thickness, with additional islands indicating the onset of additional-layer nucleation. The film is relatively flat (0.1 nm surface roughness), although there is evidence of the surface already degrading with small particles observed on the surface, due to the air exposure whilst setting up AFM. To examine the evolution of the surface roughness with exposure time, the black dashed region in Fig. 2a was measured repeatedly for 5-6 hours with each scan taking about 9 minutes. The end points of this process are shown in Fig. 2b-c, for 5 and 290 minutes of air exposure respectively. There is a clear increase in the particle coverage, and a visible change in the surface roughness as shown in Fig. 2d which plots the extracted root-mean-square (RMS) surface roughness from this region as a function of exposure time. Initially starting at 0.1 nm the roughness increases rapidly until ~200 minutes where it reaches 0.25 nm whereupon it begins to saturate and only slightly increases by 0.01 nm over the next 100 minutes. This suggests that the oxidation process does not continue to degrade the surface, and the saturated surface roughness is much less than a septuple layer thickness, this evolution is similar to other air sensitive van der Waals materials such as WTe$_2$.[13]

We now turn to synchrotron-based X-ray photoelectron spectroscopy (XPS) measurements to understand the chemical stoichiometry and thickness of the oxide layer. In Fig 3, we plot the evolution of (a) the Bi *4f* core level and (b) the Te *4d* and Mn *3p* core levels as a function of exposure time between 0 to 36 hours, taken at a photon energy of $h\nu$=350 eV. In the pristine MnBi$_2$Te$_4$ thin film, the Bi *4f* spectrum consists of two characteristic peaks representing the *4f*$_{7/2}$ and *4f*$_{5/2}$ spin-orbit split components at 157.6 eV and 163.0 eV, separated by 5.4 eV. Similarly, the Te *4d* consists of the *4d*$_{5/2}$ and *4d*$_{3/2}$ spin-orbit split components at 39.7 eV and 41.1 eV separated by 1.4 eV. The Mn *3p* core level has two components, the main Mn 3p component we assign to Mn-Te bonding within the MnBi$_2$Te$_4$ crystal structure, whilst the small shoulder at lower binding energy 47.4 eV which is marked as Mn' *3p* we assign to the partially bonded Mn ions on the surface of MnBi$_2$Te$_4$ film.[14] We attribute all these components to the characteristic bonding environment within MnBi$_2$Te$_4$ and label these peaks as green. The peak positions are consistent with previously reported values for thin-film MnBi$_2$Te$_4$.[8] In the subsequent air exposures the pristine peaks are also labelled in green.

After one hour of air exposure small oxide components are observed in the Bi *4f* core level at 159.0 eV and 164.3 eV respectively, separated from the pristine Bi *4f* by 1eV. This is in line with Bi$_2$O$_3$ previously reported for oxidized Bi$_2$Te$_3$.[10, 15] We label these Bi-O peaks as blue. The oxide to pristine component ratio Bi-O:Bi of 0.32 indicates that there is only a relatively small amount of oxidation after one hour. With increasing exposure time to 12 hours significant oxidation occurs, with the spectral weight shifting from the pristine Bi peak to the Bi-O peak, reaching a Bi-O:Bi area ratio of approximately 3. Between 12 and 36 hours of air exposure there is relatively little change in the overall peak area ratio, indicating that the surface is fully oxidized and that the oxidation process has begun to saturate. Additionally, throughout the exposure period, there is minimal shift (within experimental error) in the peak position of the pristine Bi and Te core level with air exposure, indicating a minimal change in doping of the pure MnBi$_2$Te$_4$.

In contrast to the Bi core levels, the Te *4d* core level undergoes a more significant change. After one hour of air exposure two new doublets emerge. The first doublet, at 43.6 eV and 45.1 eV, is separated from the pristine Te *4d* by 4 eV and corresponds to the formal +4 oxidation state.[16] We label this as the Te-O oxide peak in blue. The second Te *4d* doublet, separated from the pristine Te peak by 0.5 eV, appears as a shoulder to the main Te *4d* peak, and is assigned to Te atoms which are oxidized from Te$^{2-}$ in MnBi$_2$Te$_4$ to the elemental Te$^0$ state, which is plotted in purple and is consistent to previously report values.[10, 17] The main Mn *3p* core level associated with Mn-Te bonding in MBT largely remains unchanged with air exposure. A small oxide peak (Mn-O) at 51.2eV forms after 5-min exposure and increases slightly up to the 36-hour exposure. It should be noted that the partially bonded Mn disappears upon air exposure. Similar to what was observed for the Bi core level, the majority of oxidation occurs within the first 12 hours, and after that the spectra remain generally the same. It is important to note that at a photon energy of *hv* = 350 eV, the Te *4d* and Bi *4f* core levels have inelastic mean free paths (IMFP) of 9.3 Å and 7.1 Å respectively, meaning only the top surface layer is probed, with little information gained on the oxidation of the underlying MnBi$_2$Te$_4$ layers.

We determined the thickness of the oxide by performing photon-energy dependent XPS. Varying the photon energy from 350 eV to 1486 eV increases the photoelectron mean free path and allows us to probe the depth dependence of the oxidation. In Fig 4(a) we plot the Te *4d* spectra for the 36-hour air exposure at photon energies of 350 eV (upper panel), 850 eV (middle panel) and 1486 eV (lower panel). As the photon energy increases (i.e., the probing depth increases) the intensity of the oxide (blue) and elemental Te (purple) components relative to the Te components of pure MnBi$_2$Te$_4$ decreases. We have also found that the depth distribution of the elemental Te changes over time. Based on the peak area ratio, at short exposure time of 1hr, the elemental Te composition is 44.0% at photon energy of 350 eV, while the composition is 24.7% at 850 eV respectively. After long exposure of 36hrs, the composition of elemental Te decreases drastically, now being 11.4% and 15.4% respectively at 350 eV and 850 eV. This demonstrates that the formation of the oxide is a two-step process where the elemental Te is first formed and it is subsequentially oxidized to the Te$^{4+}$ state. The evolution of elemental Te composition over time suggests the formation of a Te$^0$ rich layer which moves down from the surface as the sample is oxidised. The remanent pristine peaks suggests that the oxidation is confined to the surface of MnBi$_2$Te$_4$, with little oxidation of the underlying septuple layers. Similar behavior relating to the confinement of the oxide is observed in the Bi *4f* core levels shown in Fig. 4b. The thickness of the oxide that forms on the surface of MnBi$_2$Te$_4$ can be estimated from the peak area ratio of the pristine MnBi$_2$Te$_4$ Te/Bi and the Bi-O/Te-O oxide components using the relation:[18, 19]

$$t = \lambda \ln\left(1 + \frac{A_{\text{oxide}}}{A_{\text{pristine}}}\right) \quad (1)$$

where *t* is the oxide thickness, A$_{\text{oxide}}$/A$_{\text{pristine}}$ is the ratio of the oxide and pristine peak area obtained from peak fitting, and λ is the photoelectron mean free path. We assume the mean free path is the same for MnBi$_2$Te$_4$ and the oxide and estimate this value using the NIST Electron Inelastic-Mean-Free-Path Database.[20] In Fig. 4c we plot the calculated oxide thickness as a function of exposure time. We determine the oxide thickness by taking the average thickness determined using the different photon energies and for both Bi *4f* and Te *4d* core levels. The thickness obtained from Te and Bi core levels taken at different energies using equation (1) all fall within ±5Å, and we subsequently assign a ±5Å error to the oxide thickness. The oxidation process of the thin film shows two stages: it proceeds quickly in the first 36 hours to then slow down significantly. Eventually, after even 372 hours of exposure, the oxide thickness is estimated to be still 30±5Å, consistent with the thickness of two oxidized septuple layers, given that 1SL is 1.36nm in thickness.

We now turn to measurements of the work function and valence band to understand the band alignment. Fig 5a plots the secondary electron yield vs photoelectron kinetic energy for pristine, 5 minutes, 1 hour, 12 hours and 36 hours air exposure (bottom to top). The onset in the secondary electron yield corresponds to the vacuum level of the sample with respect to the Fermi energy, and therefore directly yields the work function. For pristine $MnBi_2Te_4$, the work function is $4.40 \pm 0.05$ eV. Upon air exposure the work function reduces to a value of $4.0 \pm 0.05$ eV within 1 hour and remains unchanged (within experimental error) with further exposure, which is consistent with the general evolution of work function of metal oxide. The rapid change of work function within the first 5-min exposure can be attributed to formation of a water-related dipole layer while the further decreasing is due to the formation of the interface dipole that results in a potential step and lowers the work function.[21] We have also taken the valence band spectra at different exposure time in Fig 5b, from which $E_F$-$E_{VBM}$ can be extracted. As this is not an angle-resolved photoelectron spectroscopy (ARPES) measurement, the spectra reflect the overall joint density of states and it does not allow us to resolve the Dirac point. Instead, we are able to mark the Dirac point at 0.23 eV for the pristine film based on our recent ARPES study.[8] After air exposure, the detailed structure near the Fermi level is lost with the valence band maximum moving to around 1.6 eV below the Fermi level after 12 hours, suggesting the formation of an insulating oxide layer on top of the $MnBi_2Te_4$ thin film.

CONCLUSION

In conclusion, by combining electrical transport, AFM and XPS measurements, we have studied the dynamics of oxidation in $MnBi_2Te_4$ thin films under exposure to ambient conditions. We found that a stable few septuple layer oxide forms at the surface of $MnBi_2Te_4$ within the first several hours of exposure to air. This oxide results in only a small increase in the overall surface roughness and is confined to the top surface layers, meaning the underlying layers are protected and maintain metallic transport behavior. The demonstration of a stable oxide in $MnBi_2Te_4$ offers great potential for it to be established as a stable protective capping layer and may enable future work to develop precise control of the oxide thickness via ozone treatments, as demonstrated for other air sensitive van der Waals materials.[22]

FIGURES

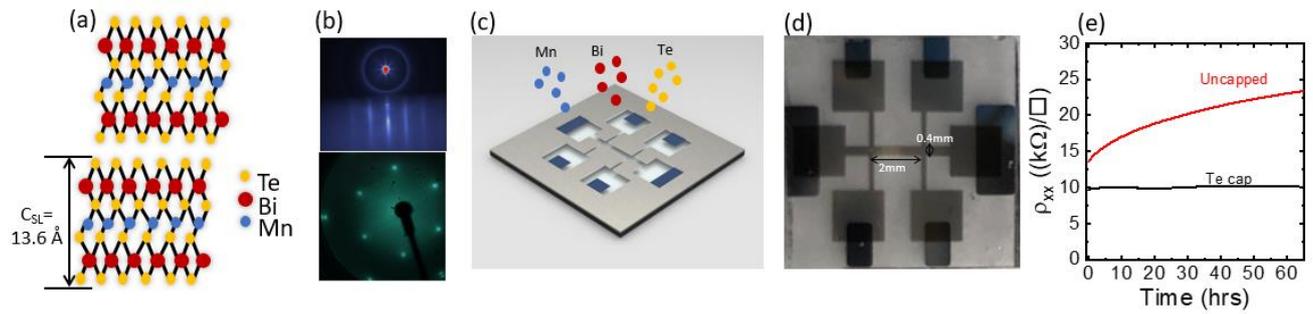

Figure 1: (a) Crystal structure of MnBi$_2$Te$_4$ consisting of Te-Bi-Te-Mn-Te-Bi-Te septuple layers. (b) Reflection High-Energy Electron Diffraction (RHEED) (upper panel) and Low energy electron diffraction (LEED) (lower panel) of 5 SL MnBi$_2$Te$_4$ grown on α-Al$_2$O$_3$(0001) and Si(111) respectively. (c) MnBi$_2$Te$_4$ growth through a shadow mask with Hall bar geometry onto an Al$_2$O$_3$ substrate with pre-deposited Pd contacts. (d) Optical image of the Hall bar device with length and width of 2mm and 0.4mm respectively. (e) Resistivity as a function of time of uncapped MnBi$_2$Te$_4$ (red) and Te capped (black) devices.

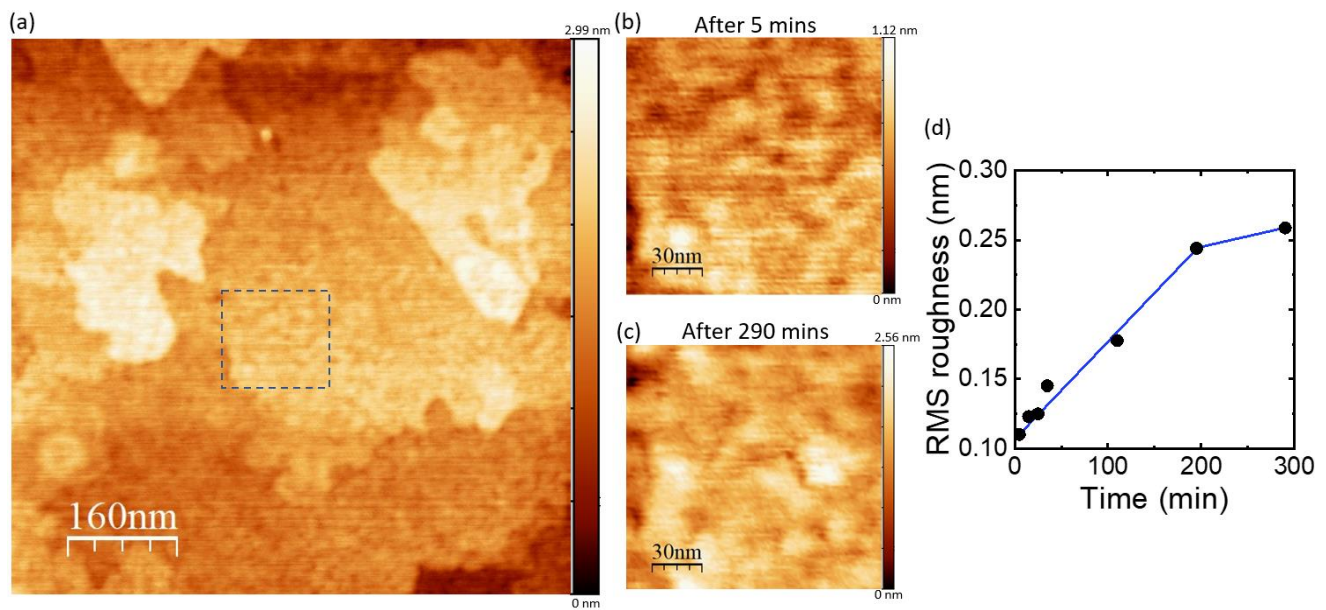

Figure 2. (a) AFM topography image (800 nm × 800 nm) of predominantly 5SL MnBi$_2$Te$_4$ on Si(111). (b) and (c) are topography images (150 nm × 150 nm) taken at the dashed box in (a) after (b) 5 min and (c) 290 min of air exposure. (d) Measured RMS roughness (extracted from the AFM images) as a function of air exposure time.

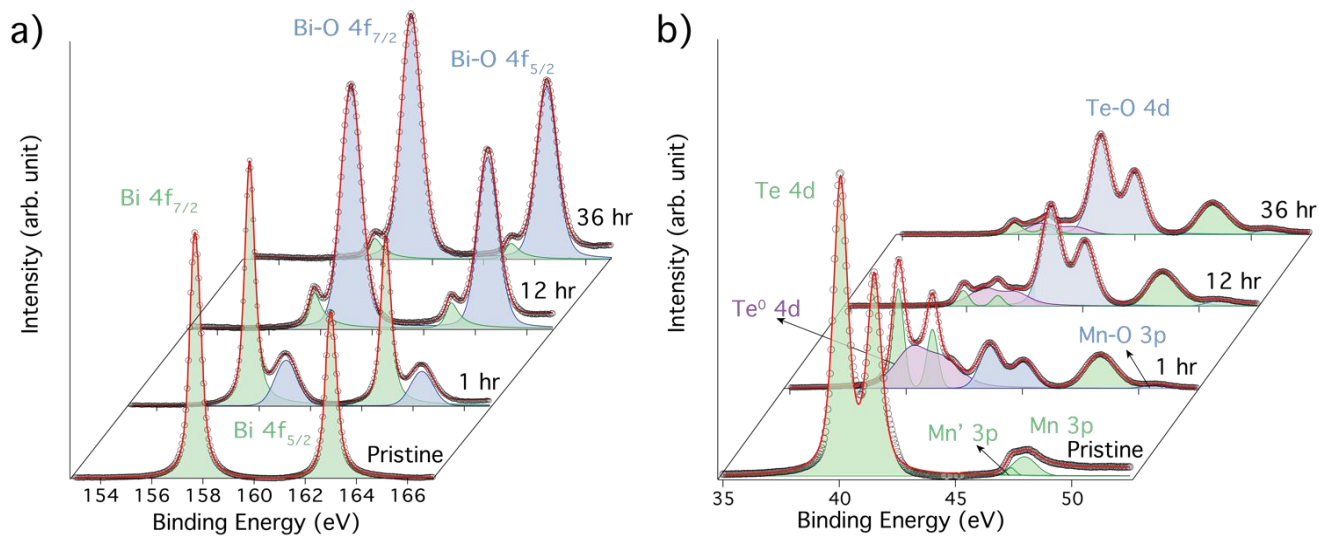

Figure 3. Evolution of 5 SL MnBi$_2$Te$_4$ after exposure to atmospheric conditions. Stack plots of (a) Bi *4f* core level and (b) Te *4d* and Mn *3p* core levels taken at *hv*= 350 eV for different exposure times between 0 and 36 hours. Experimental data is plotted as open black circles, the summed fit curve as a solid red line. Components associated with pristine MnBi$_2$Te$_4$ are shaded in green, whilst the oxide peaks are shaded in blue. There is an additional Te$^0$ state shaded in pink as described in the text.

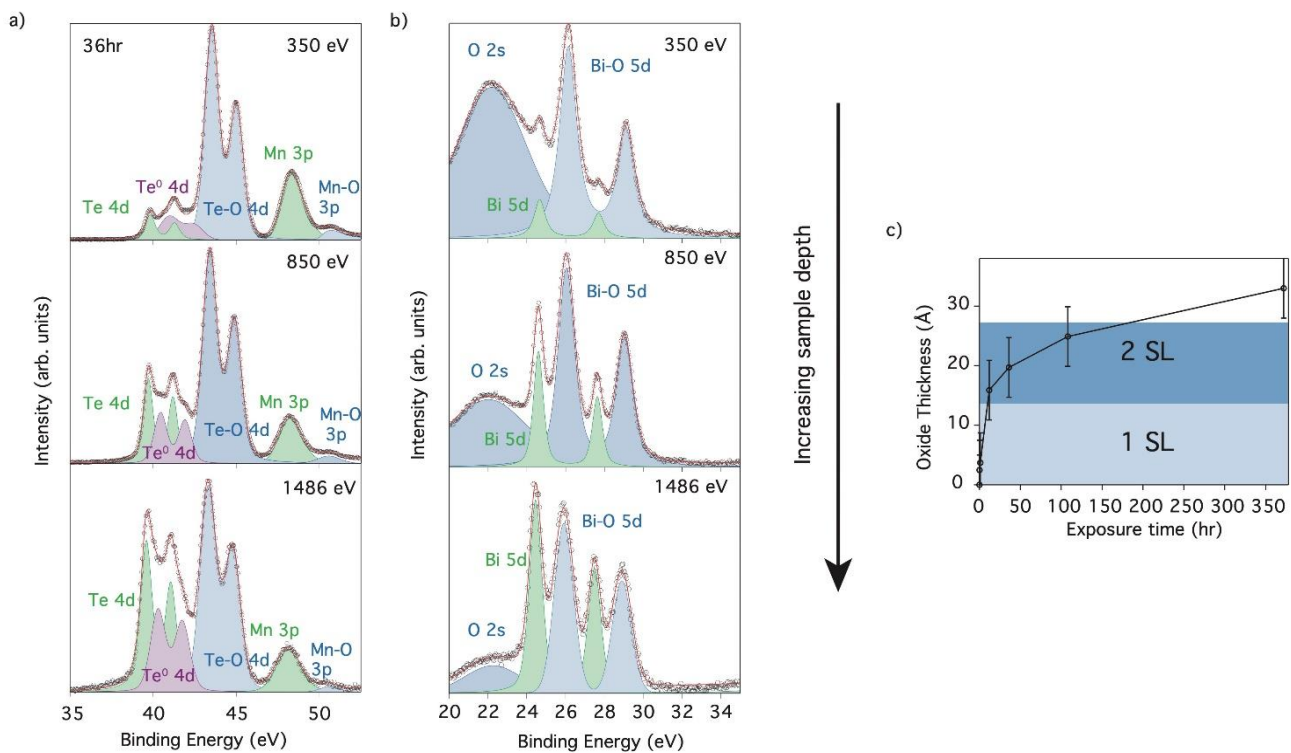

Figure 4. Depth-dependent photoelectron spectroscopy of the Te *4d* and Mn *3p* core levels (a) and Bi *5f* core levels (b) after 36hr exposure taken at $h\nu$=350 eV (upper panel), 850 eV (middle panel) and 1486 eV (lower panel). Experimental data is plotted as open black circles, the summed fit curve as a solid red line. Components associated with pristine MnBi$_2$Te$_4$ are shaded in green, whilst the oxide peaks are shaded in blue. There is an additional Te$^0$ state shaded in pink as described in the text. c) Oxide thickness as a function of air exposure time.

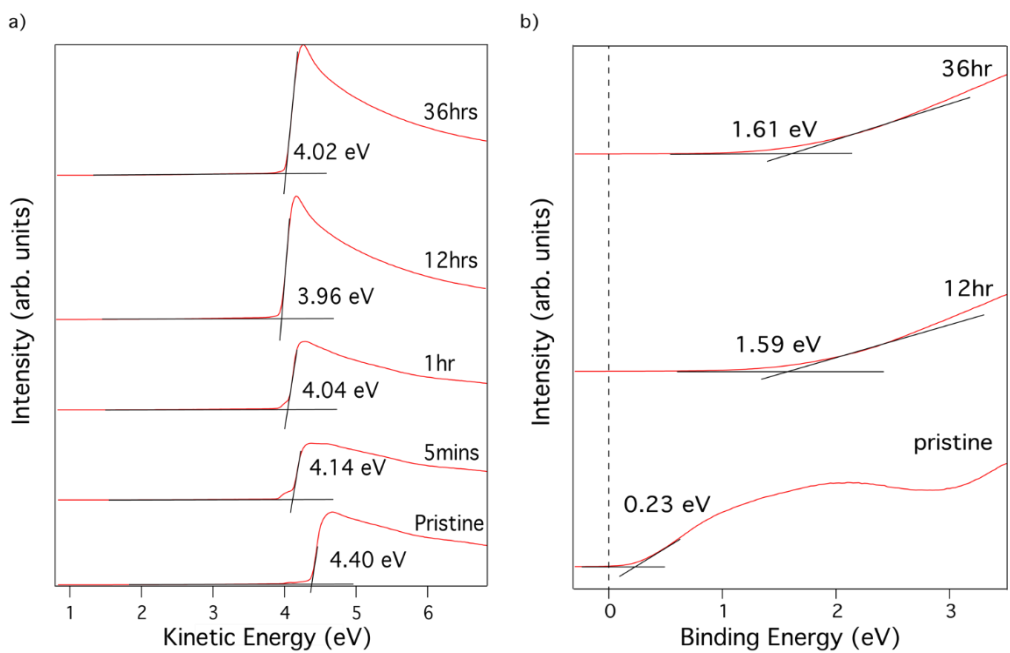

Figure 5. (a) Secondary electron cut-off (SECO) measurements showing the evolution of work function at different exposure times. The work function from SECO measurement corresponds to the energy difference of Fermi level $E_F$ and vacuum level $E_{vac}$. (b) Valence band measurements as a function of exposure time. The VB edge shifts to higher binding energy and is located 1.6 eV below the Fermi level after 36 hours, suggesting the formation of an insulating oxide layer.

ASSOCIATED CONTENT

**Corresponding Author**

*mark.edmonds@monash.edu

**Funding Sources**

Australian Research Council Centre for Excellence Future Low Energy Electronics Technologies (CE170100039). ARC DECRA fellowship (DE160101157).

**Notes**

The authors declare no competing financial interest.

ACKNOWLEDGMENT

G.A., M.T.E., J. K., and M.S.F., acknowledge the funding support from the Australian Research Council Centre for Excellence Future Low Energy Electronics Technologies (CE170100039). M.T.E. acknowledges the support from ARC DECRA fellowship (DE160101157). This work was performed in part at the Melbourne Centre for Nanofabrication (MCN) in the Victorian Node of the Australian National Fabrication Facility (ANFF). Part of this research was undertaken on the Soft X-ray beamline at the Australian Synchrotron, part of ANSTO. We would like to acknowledge Prof. Steven Prawer and used of his laser cutter facility at the University of Melbourne in making the shadow masks.

References:

1.	Li, J.; Li, Y.; Du, S.; Wang, Z.; Gu, B.-L.; Zhang, S.-C.; He, K.; Duan, W.; Xu, Y., Intrinsic magnetic topological insulators in van der Waals layered MnBi2Te4-family materials. *Science Advances* **2019,** *5* (6), eaaw5685.
2.	Otrokov, M. M.; Klimovskikh, I. I.; Bentmann, H.; Estyunin, D.; Zeugner, A.; Aliev, Z. S.; Gaß, S.; Wolter, A.; Koroleva, A.; Shikin, A. M., Prediction and observation of an antiferromagnetic topological insulator. *Nature* **2019,** *576* (7787), 416-422.
3.	Otrokov, M. M.; Rusinov, I. P.; Blanco-Rey, M.; Hoffmann, M.; Vyazovskaya, A. Y.; Eremeev, S. V.; Ernst, A.; Echenique, P. M.; Arnau, A.; Chulkov, E. V., Unique thickness-dependent properties of the van der Waals interlayer antiferromagnet MnBi 2 Te 4 films. *Physical review letters* **2019,** *122* (10), 107202.
4.	Deng, Y.; Yu, Y.; Shi, M. Z.; Guo, Z.; Xu, Z.; Wang, J.; Chen, X. H.; Zhang, Y., Quantum anomalous Hall effect in intrinsic magnetic topological insulator MnBi2Te4. *Science* **2020,** *367* (6480), 895-900.
5.	Gong, Y.; Guo, J.; Li, J.; Zhu, K.; Liao, M.; Liu, X.; Zhang, Q.; Gu, L.; Tang, L.; Feng, X., Experimental realization of an intrinsic magnetic topological insulator. *Chinese Physics Letters* **2019,** *36* (7), 076801.
6.	Liu, C.; Wang, Y.; Li, H.; Wu, Y.; Li, Y.; Li, J.; He, K.; Xu, Y.; Zhang, J.; Wang, Y., Robust axion insulator and Chern insulator phases in a two-dimensional antiferromagnetic topological insulator. *Nature materials* **2020,** *19* (5), 522-527.
7.	Zhang, D.; Shi, M.; Zhu, T.; Xing, D.; Zhang, H.; Wang, J., Topological axion states in the magnetic insulator MnBi 2 Te 4 with the quantized magnetoelectric effect. *Physical review letters* **2019,** *122* (20), 206401.
8.	Trang, C. X.; Li, Q.; Yin, Y.; Hwang, J.; Akhgar, G.; Di Bernardo, I.; Grubišić-Čabo, A.; Tadich, A.; Fuhrer, M. S.; Mo, S.-K.; Medhekar, N. V.; Edmonds, M. T., Crossover from 2D Ferromagnetic Insulator to Wide Band Gap Quantum Anomalous Hall Insulator in Ultrathin MnBi2Te4. *ACS Nano* **2021,** *15* (8), 13444-13452.
9.	Li, H.; Liu, S.; Liu, C.; Zhang, J.; Xu, Y.; Yu, R.; Wu, Y.; Zhang, Y.; Fan, S., Antiferromagnetic topological insulator MnBi 2 Te 4: Synthesis and magnetic properties. *Physical Chemistry Chemical Physics* **2020,** *22* (2), 556-563.
10.	Yashina, L. V.; Sánchez-Barriga, J.; Scholz, M. R.; Volykhov, A. A.; Sirotina, A. P.; Neudachina, V., S; Tamm, M. E.; Varykhalov, A.; Marchenko, D.; Springholz, G., Negligible surface reactivity of topological insulators Bi2Se3 and Bi2Te3 towards oxygen and water. *Acs Nano* **2013,** *7* (6), 5181-5191.
11.	Edmonds, M. T.; Hellerstedt, J. T.; Tadich, A.; Schenk, A.; O'Donnell, K. M.; Tosado, J.; Butch, N. P.; Syers, P.; Paglione, J.; Fuhrer, M. S., Stability and surface reconstruction of topological insulator Bi2Se3 on exposure to atmosphere. *The Journal of Physical Chemistry C* **2014,** *118* (35), 20413-20419.

12. Zhu, T.; Bishop, A. J.; Zhou, T.; Zhu, M.; O'Hara, D. J.; Baker, A. A.; Cheng, S.; Walko, R. C.; Repicky, J. J.; Liu, T., Synthesis, Magnetic Properties, and Electronic Structure of Magnetic Topological Insulator MnBi2Se4. *Nano Letters* **2021**.
13. Hou, F.; Zhang, D.; Sharma, P.; Singh, S.; Wu, T.; Seidel, J., Oxidation kinetics of WTe2 surfaces in different environments. *ACS Applied Electronic Materials* **2020,** *2* (7), 2196-2202.
14. Iwanowski, R.; Heinonen, M.; Janik, E., X-ray photoelectron spectra of zinc-blende MnTe. *Chemical physics letters* **2004,** *387* (1-3), 110-115.
15. Music, D.; Chang, K.; Schmidt, P.; Braun, F. N.; Heller, M.; Hermsen, S.; Pöllmann, P. J.; Schulzendorff, T.; Wagner, C., On atomic mechanisms governing the oxidation of Bi2Te3. *Journal of Physics: Condensed Matter* **2017,** *29* (48), 485705.
16. Bando, H.; Koizumi, K.; Oikawa, Y.; Daikohara, K.; Kulbachinskii, V.; Ozaki, H., The time-dependent process of oxidation of the surface of Bi2Te3 studied by x-ray photoelectron spectroscopy. *Journal of Physics: Condensed Matter* **2000,** *12* (26), 5607.
17. Sarode, P.; Rao, K.; Hegde, M.; Rao, C., Study of As2 (Se, Te) 3 glasses by X-ray absorption and photoelectron spectroscopy. *Journal of Physics C: Solid State Physics* **1979,** *12* (19), 4119.
18. Edmonds, M. T.; Tadich, A.; Carvalho, A.; Ziletti, A.; O'Donnell, K. M.; Koenig, S. P.; Coker, D. F.; Özyilmaz, B.; Neto, A. C.; Fuhrer, M., Creating a stable oxide at the surface of black phosphorus. *ACS applied materials & interfaces* **2015,** *7* (27), 14557-14562.
19. Zangwill, A., *Physics at surfaces*. Cambridge university press: 1988.
20. Powell, C. J.; Jablonski, A.; Salvat, F., NIST databases with electron elastic-scattering cross sections, inelastic mean free paths, and effective attenuation lengths. *Surface and Interface Analysis: An International Journal devoted to the development and application of techniques for the analysis of surfaces, interfaces and thin films* **2005,** *37* (11), 1068-1071.
21. Rietwyk, K. J.; Keller, D. A.; Ginsburg, A.; Barad, H. N.; Priel, M.; Majhi, K.; Yan, Z.; Tirosh, S.; Anderson, A. Y.; Ley, L., Universal work function of metal oxides exposed to air. *Advanced Materials Interfaces* **2019,** *6* (12), 1802058.
22. Kim, S.; Jung, Y.; Lee, J.-Y.; Lee, G.-H.; Kim, J., In situ thickness control of black phosphorus field-effect transistors via ozone treatment. *Nano Research* **2016,** *9* (10), 3056-3065.